\documentstyle[aps]{revtex}
\def\>{\rangle}
\def\w{\omega}
\def\W{\Omega}
\def\lr{\leftrightarrow}
\def\ud{\updownarrow}
\title{Entanglement in A Two-Identical-Particle System}
\author{Y.S. Li$^{1,2}$, B. Zeng$^{1,2}$, X.S. Liu$^3$, G.L. Long$^{1,2}$}
\address{1. Department of Physics, Tsinghua University, Beijing, 100084, China\\
         2. Key Laboratory For Quantum Information and Measurements, Beijing, 100084, 
China\\
         3. Department of Physics, Shandong Normal University, Jinan, 250014, China}
\begin{document}
\maketitle

\begin{abstract}
The definition of entanglement in identical-particle system  is introduced. The 
separability criterion in two-identical particle system is given. The physical meaning of 
the definition is analysed. Applications to two-boson and two-fermion systems are made. It 
is found new entanglement and correlation phenomena in identical-boson systems exist, and 
they may have applications in the field of quantum information.
\end{abstract}
\date{April 15, 2001}

It is no doubt that the phenomenon of quantum entanglement lies at the heart of the 
foundation of quantum mechanics. The original investigation on entanglement began from the 
famous EPR \emph{Gadanken} experiment. Entanglement has been widely applied in many 
aspects of quantum information such as quantum teleportation, quantum cryptography and 
quantum computation. However, although it is well studied in distinguishable-particle 
systems, entanglement in identical-particle system has hardly been  investigated, and  
even a proper definition is not given yet. It is noted that entanglement in certain system 
such as quantum dots\cite{r1}, Bose-Einstein condensates\cite{r2} and parametric down 
conversion\cite{r3} must be dealt with in an identical particle manner. Recently this 
problem is noted by Schliemann et al\cite{r1,r4} and they discussed the entanglement in 
two-fermion system. They have found that entanglement in two-fermion system is analogous 
to that in a two-distinguishable particle system, and the results obtained for two 
distinguishable-particle system can be translated into the two-fermion system. However due 
to the fundamental difference between bosons and fermions, the concept in two-boson system 
is quite different.  In this paper, we  explore the definition of entanglement in 
indistinguishable-particle systems. We first show that factorization alone is not able to 
define entanglement in a two-boson system. Then we gave a general definition for 
entanglement in identical particle systems. We show that this definition works well for 
two-boson system as well as for two-fermion system. Furthermore, the definition can be 
generalized into systems with more than two identical particles. We also address the 
concept of  relative correlation.

Entanglement in distinguishable-particle system has been well studied. For a system of two 
distinguishable particles possessing single particle Hilbert space labeled by ${\cal H}_1$ 
and ${\cal H}_2$, the states can be described as vectors in the direct product space 
${\cal H}_1\otimes {\cal H}_2$.
\begin{equation}
\label{e1}
|\Psi\>_{12}=\sum\limits_{i,j}c_{ij}|\phi_i\>_1\otimes|\varphi_j\>_2
\end{equation}
where $\{|\phi_i\>\}$ is basis for ${\cal H}_1$ and $\{|\varphi_j\>\}$ is basis for ${\cal 
H}_2$ respectively. The state $|\Psi\>$ is called separable if and only if it can be 
written as $|\Psi\>_{12}=|\psi\>_1\otimes|\psi^\prime\>_2$, where $|\psi\>_1\in{\cal H}_1$ 
and $|\psi^\prime\>_2\in{\cal H}_2$, otherwise it is entangled. We can define local 
operations as those operations acting on ${\cal H}_1$ and ${\cal H}_2$ respectively. A 
separable state can't be transformed into an entangled state by any local operation and 
classical communication.

 What will happen in indistinguishable-particle systems? Can the same definition be used?    
Firstly, Let's see the difference between distinguishable particle systems and identical 
systems. Suppose we have a two photon Bell state,
\begin{equation}
\label{e2}
|\lr\>_1|\ud\>_2+|\ud\>_1|\lr\>_2,
\end{equation}
where $|\lr\>$ and $|\ud\>$ stand for states with horizontal and vertical polarization 
respectively. If the two photons are separable, say the two  photons have different 
momentum though their frequencies are the same.  Then we can write state (\ref{e2}) in 
second quantization formalism as $(a_1^\dag a_3^\dag+a_2^\dag a_4^\dag)|0\>$, where 
$a_1^\dag|0\>$, $a_2^\dag|0\>$, $a_3^\dag|0\>$ and $a_4^\dag|0\>$ stands for single photon 
states $|\lr\>_1$, $|\ud\>_1$, $|\lr\>_2$ and $|\ud\>_2$ respectively. $|0\>$ is the 
vacuum state. Each photon in the system can be in one of the  four modes 
$\{a_i^\dag|0\>,\,i=1,\,2,\,3,\,4\}$, which span a 4-dimensional Hilbert space ${\cal 
H}={\cal H}_1\oplus{\cal H}_2$. $(a_1^\dag a_3^\dag+a_2^\dag a_4^\dag)|0\>$ is not 
separable and thus it is entangled. However if the two photons are indistinguishable, then 
state (\ref{e2}) will be represented by $a^{\dagger}_{\lr} a^{\dagger}_{\ud}|0\>$ which is 
separable, and hence not entangled.

 In identical particle system, it is impossible to distinguish the two particles. A direct 
sum resolution of the single particle state into two constituent particle state is not 
possible. We can only say there is one particle in a given state, but we can not tell 
which of the two particles is in that state. Because of this, separable state in identical 
particle system may be defined, analogous to the case of distinguishable particles:

 In identical two-particle system whose single particle Hilbert space ${\cal H}$ is span 
by ${\alpha_i^\dag|0\>\,i=1,\,2,\,\cdots,\,N}$, a state is separable if it can be written 
as $c^\dag d^\dag|0\>$, where $c^\dag|0\>,\,d^\dag|0\>\in{\cal H}$. Otherwise it is 
entangled.

It will be shown next that this definition does not cover all the entangled state in 
two-boson systems. The state $|\Psi\>$ of two identical bosons with single particle 
Hilbert space ${\cal H=C}^N$ can be described as follow
\begin{equation}
\label{e3}
|\Psi\>=\sum\limits_{i,j=1}^N\w_{ij}a_i^\dag a_j^\dag|0\>,
\end{equation}
where $\w_{ij}=\w_{ji}$ is an $N\times N$ complex symmetric matrix $\W$, and it can be 
decomposed to a diagonal matrix by the lemma below.

\textbf{Lemma}: For any symmetric $N\times N$ matrix $S$ there exists an unitary 
transformation $U$, such that $S=UD_MU^T$, where the matrix $D_M$ is diagonal,
\begin{equation}
D_M=\mathrm{diag}[d_1,d_2,\cdots,d_M,Z],
\end{equation}
and $Z$ is a $(N-M)\times(N-M)$ null matrix. The proof of this lemma will be given in the 
appendix.

Now, we can diagonalize the state (\ref{e3}) by a unitary transformation $U$
\begin{equation}
\label{estandard}
|\Psi\>=\sum\limits_{j=1}^M\lambda_jc_j^\dag
c_j^\dag|0\>,\hspace{1cm}c_i^\dag=\sum\limits_{j=1}^Nu_{ji}a_j^\dag,
\end{equation}
where $U$ is a representation transformation and we arrange the eigenvalues in absolute 
value descending order $|\lambda_1|\ge|\lambda_2|\ge\cdots\ge|\lambda_M|$. The     above 
diagonal form can be regarded as a standard form because it is unique except for global 
phases in the definite of two-boson basis states. Since the rank of the matrix $\W$, $M$,  
does not change under unitary transformation, and it can be used as the criterion of 
entanglement for identical two-boson system. 

\par If $N=2$, the standard  form of (\ref{estandard}) can be written as  
$(r_1e^{i\varphi}c_1^\dag c_1^\dag+r_2e^{-i\varphi}c_2^\dag c_2^\dag)|0\>$ after 
neglecting an overall phase factor. And $r_1$ and $r_2$ are nonnegtive. The state can be 
written as $[(r_1-r_2)f_1^\dag f_1^\dag+ 2\sqrt{r_1r_2} f_1^\dag f_2^\dag]|0\>$ by a 
representation transformation,
\begin{equation}
\left(\begin{array}{cc}
c_1^\dag&c_2^\dag\end{array}\right)
=\left(\begin{array}{cc}f_1^\dag&f_2^\dag\end{array}\right)
\left(\begin{array}{cc}e^{i\varphi/2}\sqrt{r_1\over 
r_1+r_2}&-ie^{i\varphi/2}\sqrt{r_2\over r_1+r_2}\\-ie^{-i\varphi/2}\sqrt{r_2\over 
r_1+r_2}&e^{-i\varphi/2}\sqrt{r_1\over r_1+r_2}\end{array}\right).
\end{equation}
\par If $r_1=r_2$, the state will be $f_1^\dag f_2^\dag|0\>$ whose rank of coefficient 
matrix is 2\cite{rlong}. It would be a separable state if definition 1 is used. \par It is 
easy to check that if (\ref{e3}) has the standard form as follows
\begin{equation}
\label{e4}
|\Psi\>=\sum\limits_{i=1}^L z_i(e^{i\varphi_i}f_{1i}^\dag 
f_{1i}^\dag+e^{-i\varphi_i}f_{2i}^\dag f_{2i}^\dag)|0\>,
\end{equation}
it can be transformed into $\sum\limits_{i=1}^L 2z_i c_{1i}^\dag c_{2i}^\dag|0\>$, which 
can be discussed as a system with distinguishable particles. If it has at least two 
nonzero $z_i$, it can be defined as distinguishable entangled state because it is 
identical to entangled states in distinguishable-particle systems. In general, a separable 
state according to definition 1 can be written as $c^\dag(\alpha c^\dag+\beta b\dag)|0\>$ 
with the rank being either 1 or 2,  where $c^\dag$ and $b^\dag$ are orthogonal. States 
such as $c^\dag(c^\dag+b^\dag)|0\>$ needs special attention. It has no invariant particle 
number in mode $c^\dag$, $b^\dag$ or $c^\dag+b^\dag$ and seems to have some persistent 
correlation. The density matrix of this state is not separable and it is an inseparable 
state. So the definition of entanglement has to be changed to let 
$c^\dag(c^\dag+b^\dag)|0\>$ be entangled and hold states such as $c^\dag d^\dag|0\>$ and 
$c^\dag c^\dag|0\>$ as separable. It is worth pointing here $c^\dag c^\dag|0\>$ can be 
regarded as if it were a single particle, and we treat them as separable. Of course, the 
definition should be generalizable to identical multi-particle systems.

Now we can give the following definition of separability and entanglement in 
identical-particle systems:\\
\textbf{Definition 2} A state with identical k-particle is separable if it can be written 
as $c_1^\dag c_2^\dag\cdots c_k^\dag|0\>$, where $c_i^\dag$ and $c_j^\dag$ are either 
equal or orthogonal. Otherwise it is entangled.

This definition works for both identical boson system and identical fermion system, 
because in the fermion system, Pauli principle prohibit two particles to occupy the same 
state, states with $c^\dag(c^\dag+b^\dag)|0\>$ $=c^\dag b^\dag |0\>$ which becomes the 
product of two orthogonal states automatically.  It is interesting to point out that an 
equivalent definition can be formulated in the following: a state with identical 
$k$-particle is separable if it is an eigenvector of a complete set of one-body hermitian 
operators, otherwise it is entangled. Complete operator sets can be generated by the 
operators $\{(a_i^\dag a_j+a_j^\dag a_i)/2,\,i,\,j=1,\,2,\,\cdots,\,n\}$. This alternative 
definition is consistent with the statement in \cite{r5,r6}, which says that any basis 
that is eigenvectors of complete set consists of one-body mechanical quantities must be 
separable in distinguishable-particle case. 

For systems with 2 bosons, we have standard form (\ref{e2}) to tell whether a state is 
entangled or not. And a state must be entangled if the rank of its coefficient matrix 
$\Omega$ is greater than 2. If $rank(\W)$ is 2, it is also easy to judge whether it is 
entangled or not from its standard form according to definition 2. From a normalized  
standard form like (\ref{estandard}), we can define the entanglement measure as
\begin{eqnarray}
-4\sum\limits_{i=1}^{[M/2]} 
(|\lambda_{2i-1}\lambda_{2i}|)\ln(|4\lambda_{2i-1}\lambda_{2i}|)
\end{eqnarray} 
for a two-boson systems. It is similar to that in distinguishable-particle systems and if 
the standard form can be written as (\ref{e4}) which has counterpart in distinguishable 
system, the entanglement measure thus defined will be just the partial entropy in 
distinguishable-particle systems\cite{entropy}. But for systems with $k$k($k>2$) bosons, 
it is more difficult to give the standard forms to tell whether a state is separable. It 
is also noted that the    quantitative description  of  entanglement in multi-particle 
systems is a very hard mathematical problem, and it is still not solved in 
distinguishable-particle systems. The problem to quantify entanglement in identical-boson 
systems is even more difficult and remains an    open challenge. 

New entanglement has been found in identical boson systems. They may have important  
applications different from those in distinguishable-particle systems. For example 
$(a^\dag a^\dag+b^\dag c^\dag)|0\>$ is an entangled state in an identical boson system 
with $N=3$ single particle space. It  can denote a superposed state of two photons states, 
one with two photons sent to Alice, and one with one photon sent to Bob and the other sent 
to Clare. If Alice, Bob and Clare measure photon number  respectively. If Alice gets two 
photons, none photon will reach Bob and Clare, and if no photon reach Alice, either of Bob 
and Clare will have one photon. If Bob get one photon, the other photon will reach Clare. 
Situation of Clare is similar to Bob. So if one person of the three get a result, the 
results of the other two are decided, which means that some kind of communication can be 
built with only two photons.

The above definition of entanglement also applies to many identical fermion system.  
Schliemann et al have discussed entanglement in two-fermion systems\cite{r4}. For a 
two-fermion system having the single particle Hilbert space ${\cal C}^{2K}$, an arbitrary 
state has the form
$|\Psi\>=\sum\limits_{i,j=1}^{2K}\w_{ij}a_i^\dag a_j^\dag|0\>$ ,where $\W=(\w_{ij})$ are 
anti\-symmetric. It can be decomposed into a standard form 
$\sum\limits_i^K z_if_{1i}^\dag f_{2i}^\dag|0\>$ by a representation 
transformation\cite{r4}, which implies that entanglement and separability in two-fermion 
systems are equivalent with those in distinguishable-particle systems. However when 
generalizing this into many fermion system, this elegant property in two-fermion system 
disappear in many fermion system. This can be easily understood that the single particle 
Hilbert space ${\cal H}$ for a state in system with $k$ greater than 2 identical  fermions 
can't  be decomposed to a direct sum resolution of k subspaces.     For example, composing 
a three-fermion state with single particle Hilbert space ${\cal C}^{3N}$ to three 
orthogonal N-dimension subspaces requires  $7N^3-9N^2+2N$ real equations satisfied while 
the group $SU(3N)$ can only provide $9N^2-1$ real parameters. The equations have the 
required number of parameters only when $N=1$ or $N=2$. It is easy to check the definition 
1 and definition 2 are equivalent for identical fermion system because arbitrary two 
fermions can't be in one state. Hence for identical two-fermion systems, the rank of $\W$ 
is the  criterion to judge entanglement or separability: $rank(\W)=2$ for separability and 
$rank(\W)>2$ for entanglement. It must be noted that $rank(\W)\neq 1$ for two-fermion 
systems.

Another important concept is quantum correlation. It is quite often used in literatures of 
physics. A recent development has made this concept an important one. It is well known 
that quantum teleportation can be implemented with Bell states which are distinguishably 
entangled. Lee et al gave an experimental scheme, in which a state superposed by one 
photon and vacuum can be teleported with a single photon state 
$(a^\dag+b^\dag)|0\>$\cite{r7}, where $a^\dag$ and $b^\dag$ are particle creation 
operators in path A and B. It is meaningless to discuss entanglement for a single 
particle, therefore there is not entanglement in this teleportation scheme. They suggests 
that entanglement may be not necessary for quantum teleportation, because it can even be 
implemented with separable states. To study phenomenon like this, it is useful to define 
relative correlation:

\textbf{Definition 3} A state is said to have correlation relative to a quantum-mechanical 
quantity $F$, if and only if the state is not an eigenvector of $F$.

According to the definition above,  the quantity $F$ is important for certain correlation. 
A correlation must be related to certain measurement corresponding to the mechanical 
quantity $F$. In fact, the so-called correlation is the correlation between eigenvectors 
with different eigenvalues of $F$. Operations in eigenvectors with the same eigenvalue may 
be called local operations. Operations in eigenvectors with different eigenvalues are 
nonlocal. It is obvious that local operations do not change eigenvectors' eigenvalues. 
There is correlation relative to particle number $a^\dag a$ or $b^\dag b$ in 
$(a^\dag+b^\dag)|0\>$ and no correlation relative to the two particle number operators in 
$a^\dag b^\dag|0\>$, where $a^\dag$ and $b^\dag$ are orthogonal. For the state $a^\dag 
b^\dag|0\>$, local operations relative to $a^\dag a$ do not affect the particle at mode 
$b^\dag$, and the similar occurs to $b^\dag b$. Hence $a^\dag b^\dag|0\>$ is called a 
separated state in distinguishable-particle systems. Entanglement must have some 
correlation,  but correlation can happen in both separable and entangled states.  

To summarize, entanglement in identical particle system can be well distinghuished by 
definition 2. This definition reduces to that in distinguishable-particle system if the 
particles are distinguishable. It is noted that    the entanglement definition of both  
distinguishable-particle and identical-particle systems  can be dealt with using the 
definition given in this paper. Using identical particle formalism to treat identical 
particle system  is important,  examples of such treatment can be found in Refs. 
\cite{r8,r9}. There are also new phenomena in identical-particle systems, which may have 
future applications. Moreover, relative correlation is defined, and it may be  physically 
more important  than entanglement.
\section*{Proof of Lemma}
\emph{Proof}: Let $S$ be a $N\times N$, complex, symmetric matrix, $S^T=S$. Hence 
$SS^*=SS^\dag$ is hermitian which can be diagonalized by a unitary transformation 
$U^\prime$:
$SS^*=U^\prime DU^{\prime\dag}$, D - diagonal. Let us define
$C:=U^{\prime\dag}SU^{\prime*}$. It is easy check that C is symmetric and normal 
$CC^\dag=C^\dag C$. Then we'll decompose $C$ into its real and imaginary parts: $C=F+iG$. 
Since C is normal, F and G commute. Thus $F$ and $G$ are real, symmetric, commuting 
matrices. Hence they can be simultaneously diagonalized by a real orthogonal 
transformation $O$, $F=OD^1O^T$ and $G=OD^2O^T$. Thus $C=OD_MO^T(D_M=D^1+D^2)$ and finally
\begin{equation}
S=U^\prime OD_MO^TU^{\prime T}=UD_MU^T,
\end{equation}
where $U=U^\prime O$.
\section*{ACKNOWLEDGEMENT}
The authors would like to thank Prof. J.Y. Zeng, Prof. C.P. Sun, Prof. S.Y. Pei, Mr. P. 
Zhang and Mr. H. Zhai for useful discussions. This work is supported by the China National 
Natural Science Foundation Grant No. 60073009, the Fok Ying Tung Education Foundation, and 
the Excellent Young University Teachers' Fund of Education Ministry of China. 


\begin{thebibliography}{99}
\bibitem{r1} J. Schliemann, D. Loss and A.H. MacDonald, Phys. Rev. B63(2001)085311
\bibitem{r2} A. S$\phi$rensen, L.M. Duan, J.I. Cirac and P. Zoller, Natrue 409(2001)63.
\bibitem{r3} Y.H. Kim, M.V. Chekhova, S.D. Knlik, M.H. Rubin and Y. Shih, to appear in 
Phys. Rev. A. Also available in quant-ph/0103168.
\bibitem{r4} J. Schliemann, J.. Cirac, M. Ku$\acute{u}$s, M. Lewenstein and D. Loss, 
quant-ph/0012094.
\bibitem{rlong} $f_1^\dag f_2^\dag|0\>=\sum_{i,j}\w_{i,j}f^{\dag}_i f^{\dag}_j=(f_1^\dag 
f_2^\dag|0\>+f_2^\dag f_1^\dag|0\>)/2$ 
\bibitem{r5} H.B. Zhu and J.Y. Zeng, to be published in Science in China.
\bibitem{r6} H.B. Zhu and J.Y. Zeng, to be published in Chin. Phys. Lett.
\bibitem{r7} H.W. Lee and J. Kim, Phys. Rev. A.63(2000)012305
\bibitem{r8} J. Calsamiglia and N. L$\ddot{u}$tkenhaus, Appl. Phys. {\bf B 72}(2001) 67. 
Also in quant-ph/0007058.
\bibitem{r9} N. L$\ddot{u}$tkenhaus, J. Calsamiglia and K.A. Suominen, Phys. Rev. 
A59(1999)3295.
\bibitem{entropy} C.H. Bennett, D.P. Divincenzo, J.A. Smolin and W.K. Wootters, Phys. Rev. 
A54(1996)3824, quant-ph/9604024
\end{thebibliography}
\end{document}